\documentclass[onecollarge,natbib]{svjour2}
\bibpunct{[}{]}{;}{n}{}{,} 
\smartqed  
\usepackage{graphicx}
\journalname{Few-Body Systems}
\begin{document}

\title{Gluon and Ghost Dynamics from Lattice QCD
}


\author{O.~Oliveira \and A.~G.~Duarte \and D.~Dudal \and P.~J.~Silva}


\institute{O. Oliveira, A. G. Duarte, P. J. Silva \at
               CFisUC, Department of Physics, University of Coimbra P-3004-516 Coimbra, Portugal \\
              \email{orlando@fis.uc.pt, agd077@hotmail.com, psilva@uc.pt}           
           \and
           D. Dudal \at
           KU Leuven Campus Kortrijk - KULAK, Department of Physics, Etienne Sabbelaan 51 bus 7800, 8500 Kortrijk, Belgium and \at Ghent University, Department of Physics and Astronomy, Krijgslaan 281-S9, 9000 Gent, Belgium    \\
           \email{david.dudal@kuleuven.be}       
}

\date{Received: date / Accepted: date}

\maketitle

\begin{abstract}
The two point gluon and ghost correlation functions and the three gluon vertex are investigated, in the Landau gauge, using lattice simulations.
For the two point functions, we discuss the approach to the continuum limit looking at the dependence on the lattice spacing and volume.
The analytical structure of the propagators is also investigated by computing the corresponding spectral functions using an implementation of the
Tikhonov regularisation to solve the integral equation.
For the three point function we report results when the momentum of one of the gluon lines is set to zero and discuss its implications.
\end{abstract}

\section{Introduction and Motivation}
\label{intro}

In this proceeding we report on ongoing work concerning the computation of the Landau gauge two and three point functions in pure Yang-Mills SU(3)
theory using lattice QCD methods.

The gluon propagator started to be investigated within the lattice QCD approach in the late eighties of the last century~\cite{Mandula87}.
At the time, the focus was on the nature of the propagator and in particular if the gluon behaves as massive particle. In the nineties, there was
a renewed interest due to various proposals for the non-perturbative quantisation of the Yang-Mills theory, some of which made precise predictions
for the behaviour of gluon and ghost propagators at zero momentum.

In what concerns the evaluation of the ghost propagator using lattice QCD methods, a first calculation was reported only during the 1990s~\cite{Suman96}.

On the other hand, in the late nineties of the last and by the beginning of this century, the evolution of the continuum approach to QCD
based on the Dyson-Schwinger equations provided another non-perturbative approach which, together with lattice QCD and the increase in
computer power, enabled a better comprehension of the fundamental QCD Green functions. Furthermore, new developments on the theoretical side
and new numerical simulations gave an important contribution towards our current understanding of the propagators and, partially, on higher
order Green functions of QCD.

For the gluon propagator it is now clear that it is finite over the whole range of momenta and it does not vanish at $p = 0$. The non-linear dynamics of QCD
is translated into a gluon running mass whose value is about $\sim 2 \, \Lambda_{QCD}$ at zero momentum. For the ghost propagator, lattice QCD
methods and continuum methods provided essentially the same results and suggest that the propagator follows closely the behavior of a free massless
(ghost) particle. 
Useful reference works on the analytical study of QCD Green's functions are  \cite{Alkofer:2000wg,Dudal:2008sp,binosi,Boucaud:2011ug,Bashir:2012fs}.

The ghost propagator being essentially the propagator of massless particle has profound implications for the Green functions with higher number of
gluon legs. Indeed, if one requires that the renormalised Dyson-Schwinger equations are finite, then the full three gluon vertex should change sign
for momenta $\sim 200$ MeV \cite{Binosi13}.
On the other hand, the vanishing of the three gluon vertex also implies that the gluon propagator should have a local
maximum at about the same momenta. According to the predictions of the Dyson-Schwinger equations this maximum is very mild and, therefore,
it will be very hard to observe in lattice simulations. Anyway, if our current understanding of the QCD dynamics is not that far from the
``true  exact dynamics''
one expects to observe, at least, a change in sign of the three gluon vertex.

Our motivation to revisit the calculation of the gluon and ghost propagators is twofold. Previous simulations which helped to improve the understanding
of these two point functions were done on large lattices but using the quite large lattice spacings of the order
$\sim 0.2$ fm~\cite{Cucchieri:2007md,Bogolubsky:2009dc}.
 Also the approach to the
continuum limit relied on relative large lattice spacings. Therefore, it is important to check the results of the previous simulations and also to understand
how the results of the lattice simulations approach the continuum. Having access to the two propagators, we can also check how the strong coupling
changes with the lattice spacing and the lattice volume. We call the reader's  attention to the observation that what we call here the strong coupling constant is
a particular combination of the gluon and ghost propagators which is a renormalisation group invariant quantity but other definitions for $\alpha_s(p^2)$
are possible. Further, from the gluon and ghost propagators one can also investigate the so-called spectral representation via a K\"all\'en-Lehmann
representation. The computation of the spectral representation requires inverting an integral equation and this can be achieved via the so-called
 (linear) Tikhonov regularisation. The spectral representation for the gluon and the ghost can provide us information about the nature of these
fields and, in particular, if the Hilbert space of QCD $\mathcal{H}$ includes single gluon and/or ghost particle states. If such states are not within
$\mathcal{H}$, then gluons and ghosts can only belong to the physical states as constituents of other particles and, in this sense, are confined.

The three gluon vertex is an important element of the continuum approach to QCD and it provides an important test to the full tower of Green functions
and, in particular, a test of the nature of the ghost propagator. Our final aim is to provide a complete as possible description of the three gluon vertex but herein
we just report results for a single kinematical configuration.

The current proceeding summarises the results reported before
 in~\cite{Oliveira12,Duarte16a,Duarte16b} and the interested reader should consult these
references for further details. Here we also report on our investigation of the gluon and ghost spectral density computed as described in~\cite{Dudal14}.

\section{The Gluon Propagator}
\label{sec:1}
In Euclidean space and in the Landau gauge the gluon and ghost propagators are given by
\begin{equation}
D^{ab}_{\mu\nu} (p) = \delta^{ab}\left( \delta_{\mu\nu} - \frac{p_\mu p_\nu}{p^2}\right) D(p^2)
\qquad\mbox{ and }\qquad
G^{ab} =  - \delta^{ab} G(p^2) \ ,
\end{equation}
where the latin indices mean color degrees of freedom and the greek indices Lorentz degrees of freedom.

\begin{figure}
\centering
  \includegraphics[width=0.45\textwidth]{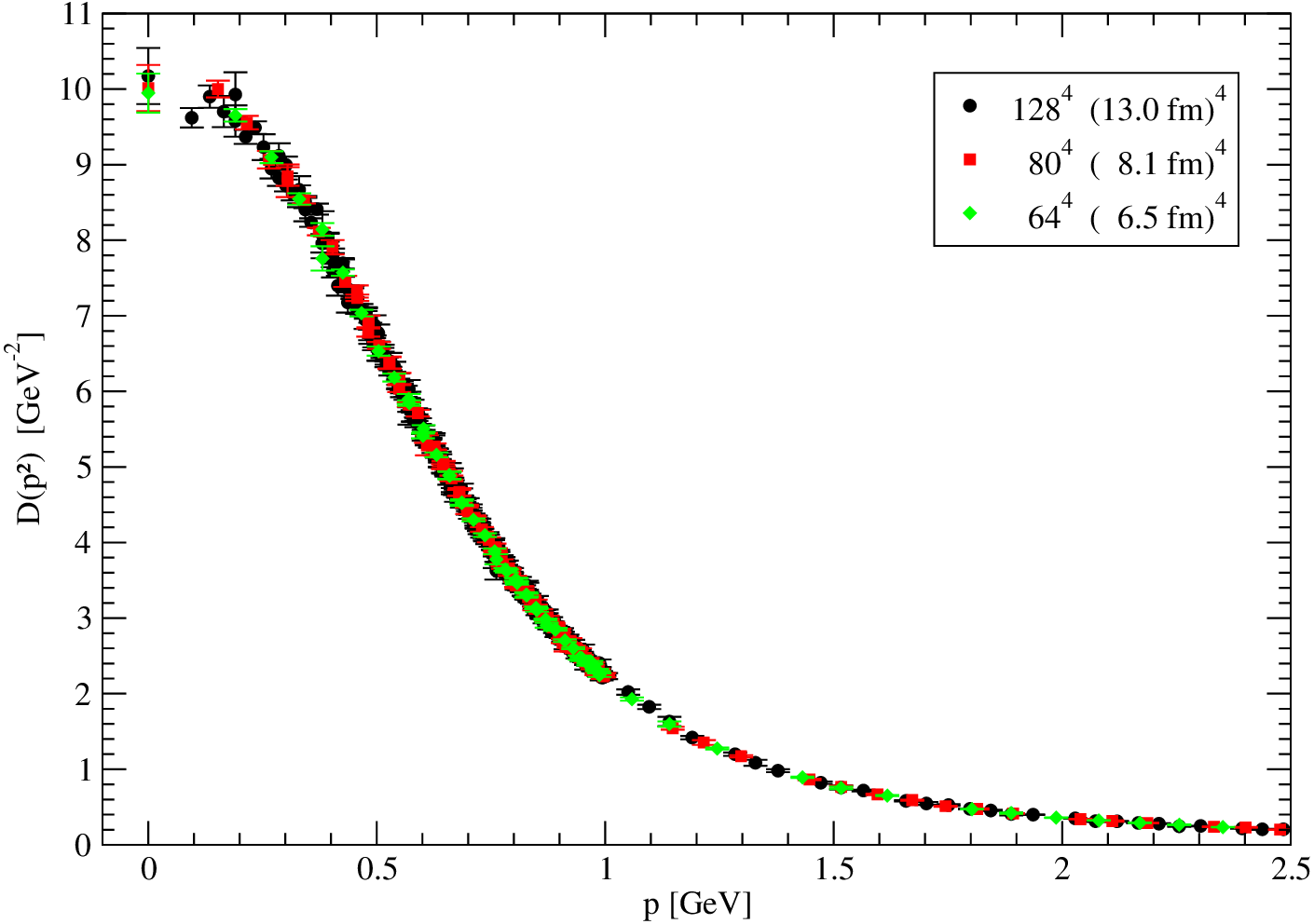} \quad
  \includegraphics[width=0.45\textwidth]{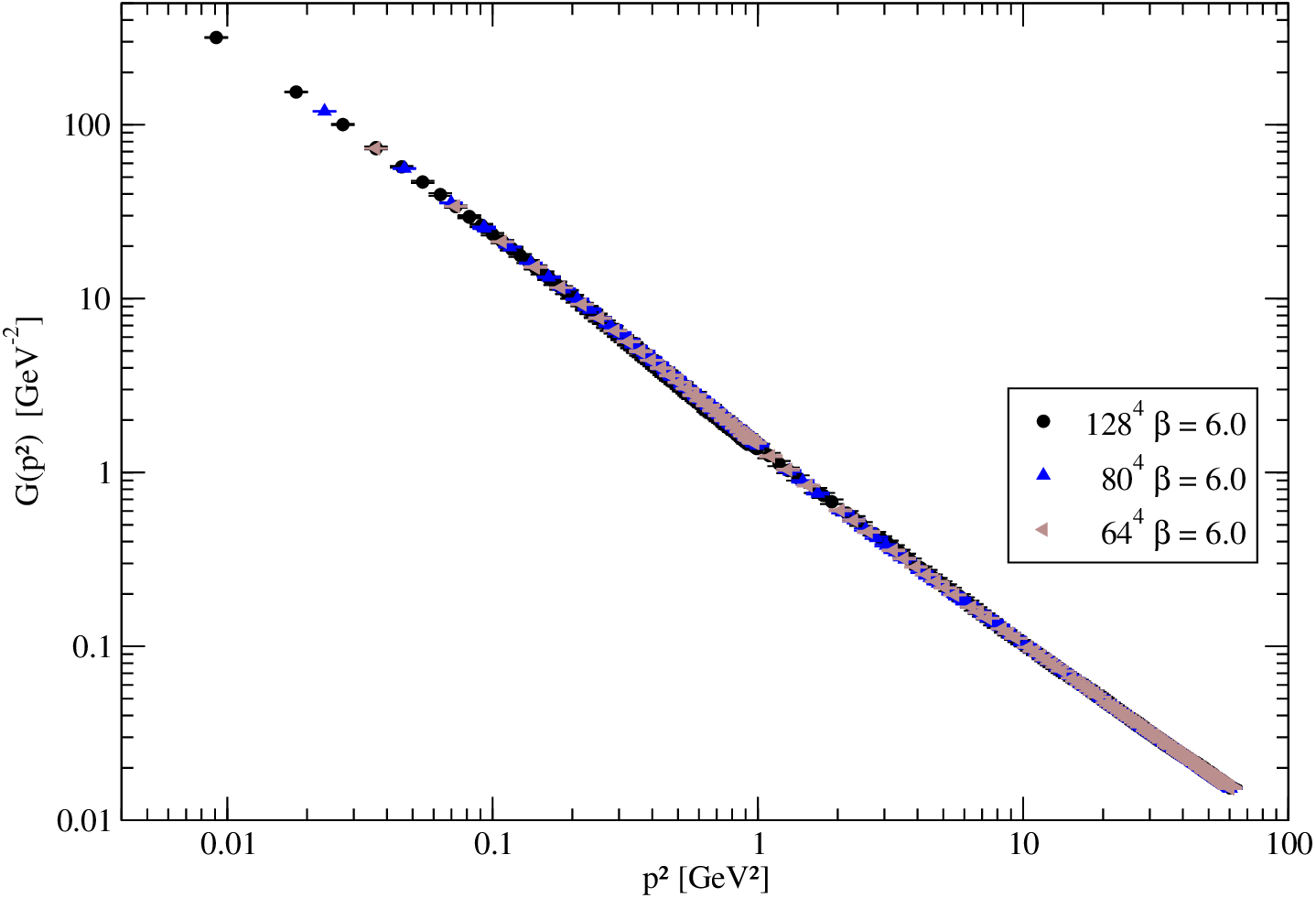}
\caption{Gluon (left) and ghost (right) propagator compute for $\beta = 6.0$ and various lattice volumes and renormalized at $\mu = 4$ GeV.
The lattice spacing associated with these simulations being $a = 0.1016$ fm.}
\label{fig:1}       
\end{figure}

The general features of the gluon and ghost propagators can be seen on Fig.~\ref{fig:1}. The gluon propagator in the infrared region can
be described by a Yukawa like propagator~\cite{Oliveira2011} and, therefore, its infrared behaviour can be approximated by a quasi-particle
massive type of propagator. In the ultraviolet region the lattice data is well described by one-loop renormalisation group improved perturbation theory.
Moreover, as a function of $p^2$, the function $D(p^2)$ shows two different curvatures which can be linked to different signs for its
spectral density. The different curvatures can be interpreted as a sign of positivity violation and, therefore, $\mathcal{H}$ should not contain one gluon
states. On the other side, as the figure shows the ghost propagator retains essentially its perturbative form and, in this sense, behaves as a massless
particle. Indeed, we have tried to fit $G(p^2)$ to the one-loop renormalisation group improved perturbative expression and verified that the lattice
data and perturbation theory are already compatible for momenta above $\sim 800$ MeV.

\begin{figure}
\centering
  \includegraphics[width=0.45\textwidth]{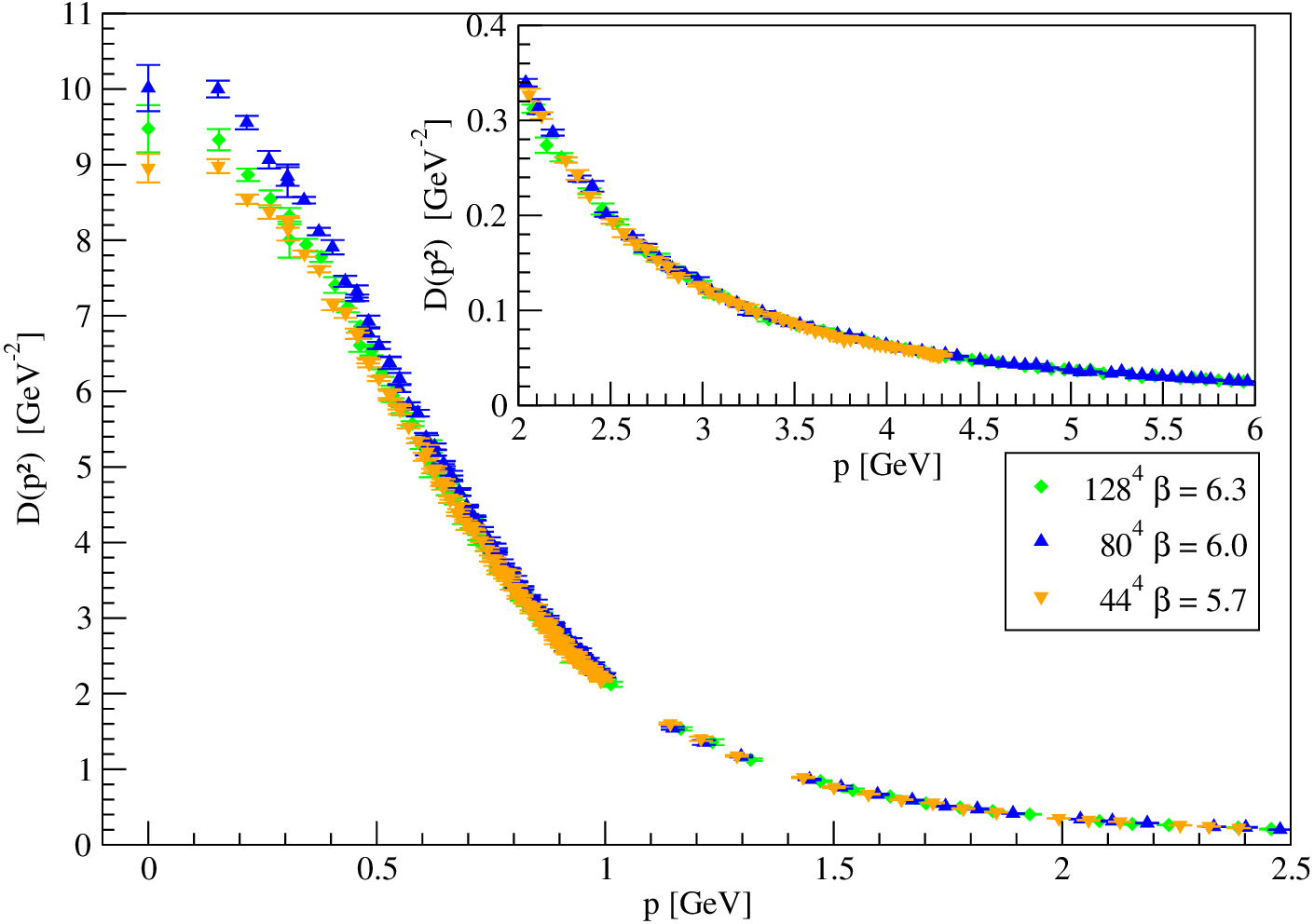} \quad
  \includegraphics[width=0.45\textwidth]{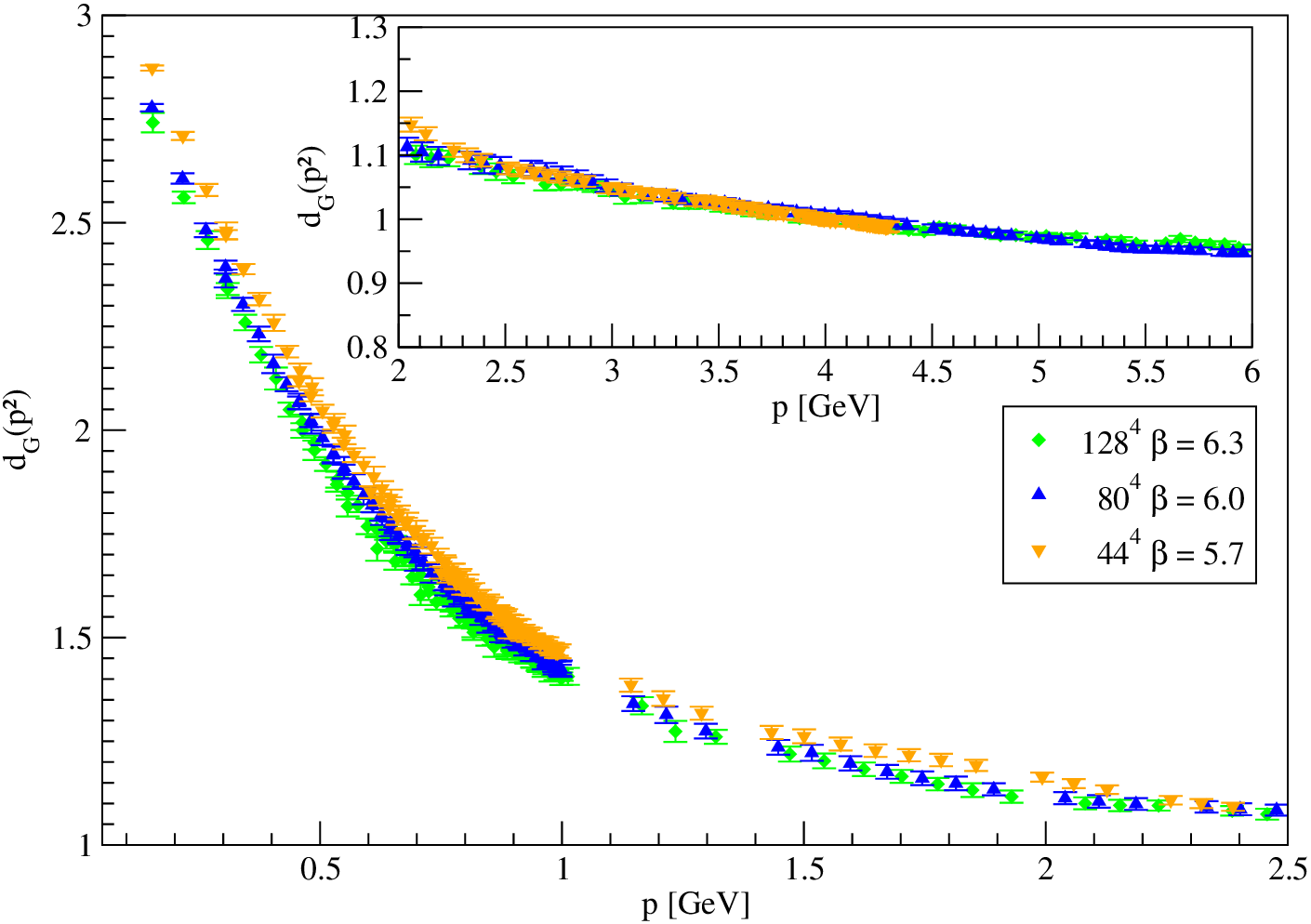}
\caption{Gluon propagator (left) and ghost dressing function $d_G(p^2) = p^2 G(p^2)$ (right) compute using different lattice spacings
and the same physical volume $V \sim ($8 fm$)^4$. All data is renormalized at $\mu = 4$ GeV.
The lattice spacing associated with these simulations being $a = 0.1838$ fm for $\beta = 5.7$, $a = 0.1016$ fm for $\beta = 6.0$ and
$a = 0.0627$ for $\beta = 6.3$.}
\label{fig:2}       
\end{figure}

The Fig.~\ref{fig:1} also shows that the volume dependence of the lattice results for lattices with physical volumes greater than
(6.5 fm)$^4$ is very mild.

The dependence of the lattice propagators with the lattice spacing, for the same physical volume, is not trivial and can be seen in Fig.~\ref{fig:2}.
For the gluon propagator the differences show up for momenta smaller than $\sim 500$ MeV and, as the figure shows, the dependence is not
monotonous. The simulation using the larger value for the lattice spacing lies always below all the remaining data. On the other hand,
the data for the ghost propagator points towards a decrease of the lattice propagator as one approaches the continuum limit. For the ghost, the
differences between the various sets of data appear at all momenta. Our simulations suggest that the results for $\beta = 5.7$ provide lower bounds
for the gluon propagator and upper bounds for the ghost propagator.

\begin{figure}
\centering
  \includegraphics[width=0.45\textwidth]{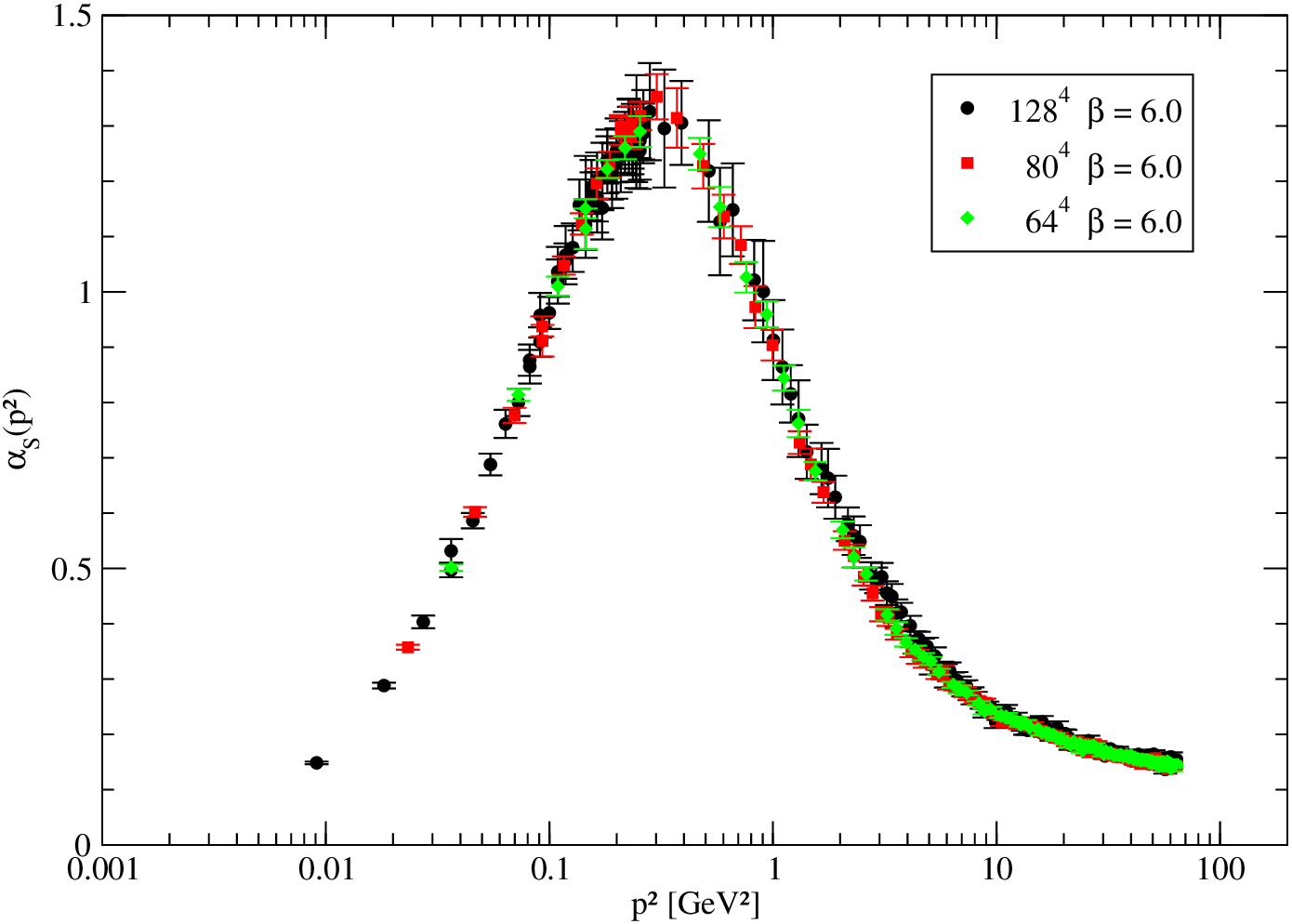} \quad
  \includegraphics[width=0.45\textwidth]{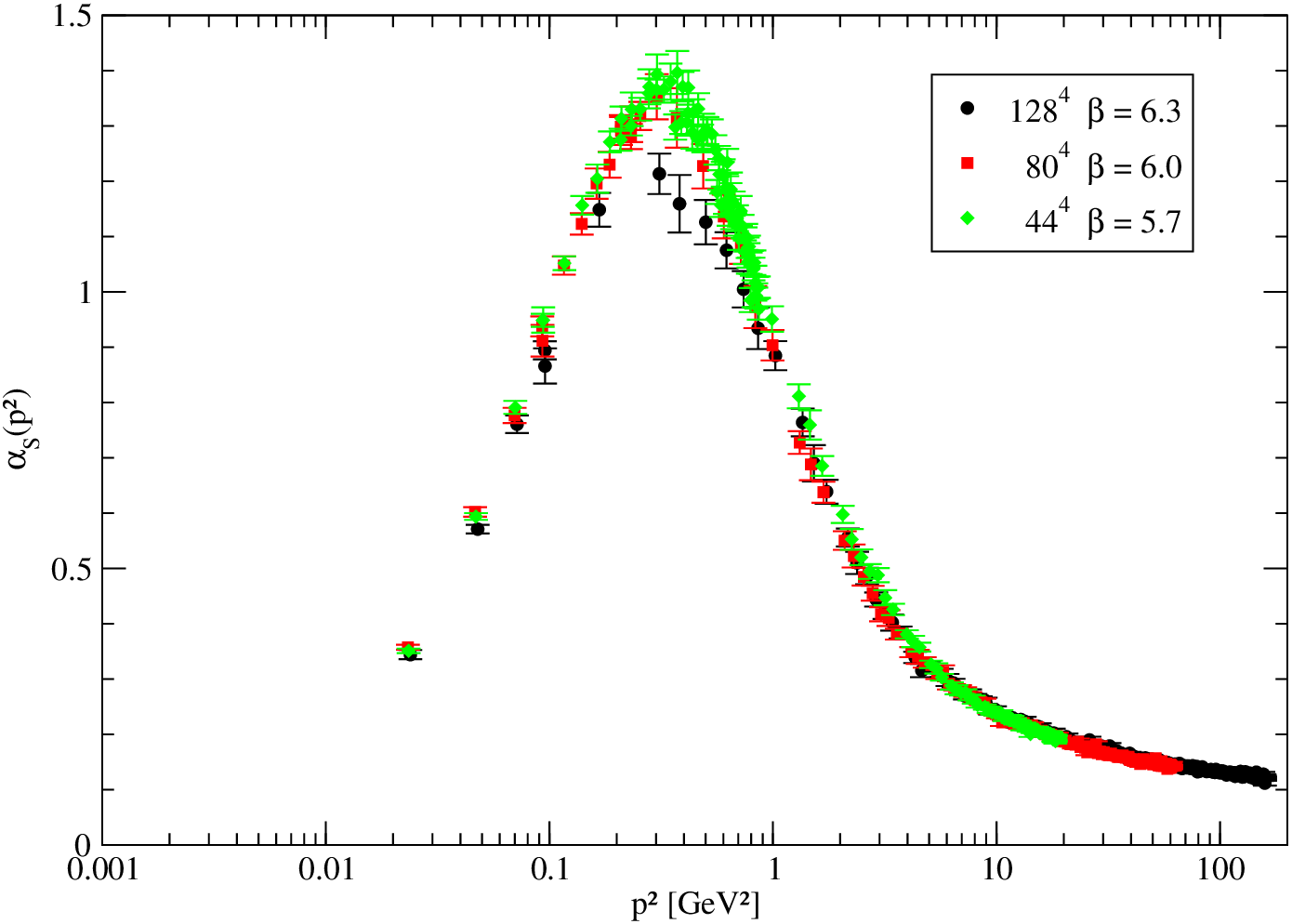}
\caption{Volume dependence (left) and lattice spacing dependence (right) of the strong coupling constant as defined by
Eq. (\ref{Eq:alpha}). For the values of the lattice spacing see caption of Fig.~\ref{fig:2}.}
\label{fig:3}       
\end{figure}

In the Landau gauge one can define a renormalisation group invariant strong coupling constant from the gluon and ghost propagators.
The definition is\footnote{This is one of the many possibilities to define a strong coupling constant over the whole momentum range, see the discussions in e.g.~\cite{Lerche:2002ep,Aguilar:2009nf,Binosi:2016nme}, which all share the same property of having the desired UV perturbative behaviour. The upshot of using the specific definition of (\ref{Eq:alpha}) is that this can be accessed from lattice data.}
\begin{equation}
   \alpha_s (p^2) = \frac{g^2}{4 \pi}  ~ d_D(p^2) \  d^2_G(p^2) \ ,
   \label{Eq:alpha}
\end{equation}
where $d_D (p^2) = p^2 D(p^2)$ and $d_G (p^2) = p^2 G(p^2)$ are the gluon and ghost dressing functions, respectively,
and the lattice data can be seen in Fig.~\ref{fig:3}. As the Fig. shows, $\alpha_s(p^2)$ has a mild or almost no dependence on the physical
volume used in the simulation. However, the strong coupling constant seems to have a smaller maximum, although its position occurs
at the same momentum, when the lattice spacing is decreased.

We would like to call the reader's attention that, although we see nontrivial dependences of the lattice results with the lattice spacing,
the functional forms of the propagators and the strong coupling constant seem to be same. Indeed, the lattice propagators and, therefore,
the strong coupling are well described by the same function forms.

\section{Gluon and Ghost Spectral Representation}

The K\"all\'en-Lehmann representation associated to the gluon propagator is given by
\begin{equation}
   D(p^2) = \int^{+ \infty}_0 d \mu ~ \frac{\rho(\mu)}{p^2  + \mu} \ ,
   \label{Eq:spectral}
\end{equation}
where the spectral function $\rho ( \mu )$ would be a positive definite function if the Hilbert space of states of QCD includes asymptotical one gluon states.
If $\mathcal{H}$ does not include such gluon states, then it is not clear if the spectral integral (\ref{Eq:spectral}) even holds to begin with.
For the remainder of this proceeding, we will nevertheless assume
that the gluon and ghost propagators have an associated spectral function defined in the usual way
and we shall try to get some information on the
function $\rho ( \mu )$ from the lattice data. A non-positive spectral function will be interpreted as a sign that gluons are confined, i.e. can
only be part of composite states of $\mathcal{H}$.

According to renormalization group 
improved perturbation theory (leading log resummation), for sufficiently high momenta,
the gluon propagator is given by
\begin{equation}
   D(p^2) \propto \frac{1}{p^2} \left[ \frac{\ln \frac{p^2}{\Lambda_R^2}}{\ln \frac{\mu_R^2}{\Lambda_R^2}} \right]^{-\gamma} \ ,
\end{equation}
where $\mu_R$ is the renormalisation scale, and $\Lambda_R$ the renormalization group invariant QCD scale in a given scheme $R$, here we have $\gamma = 13/22$, up to the sign, the gluon anomalous dimension rescaled by the leading order $\beta$-function coefficient. Given that $D(p^2)$ decreases
faster than $1/p^2$ at high momenta, then the spectral density satisfies the following sum rule
\begin{equation}
  \int d \mu ~ \rho ( \mu ) = 0 \ ;
  \label{Eq:sumrule}
\end{equation}
see \cite{Cornwall13} for a review on spectral representations for non-Abelian gauge theories and for some further details.

\begin{figure}
\centering
  \includegraphics[width=0.45\textwidth]{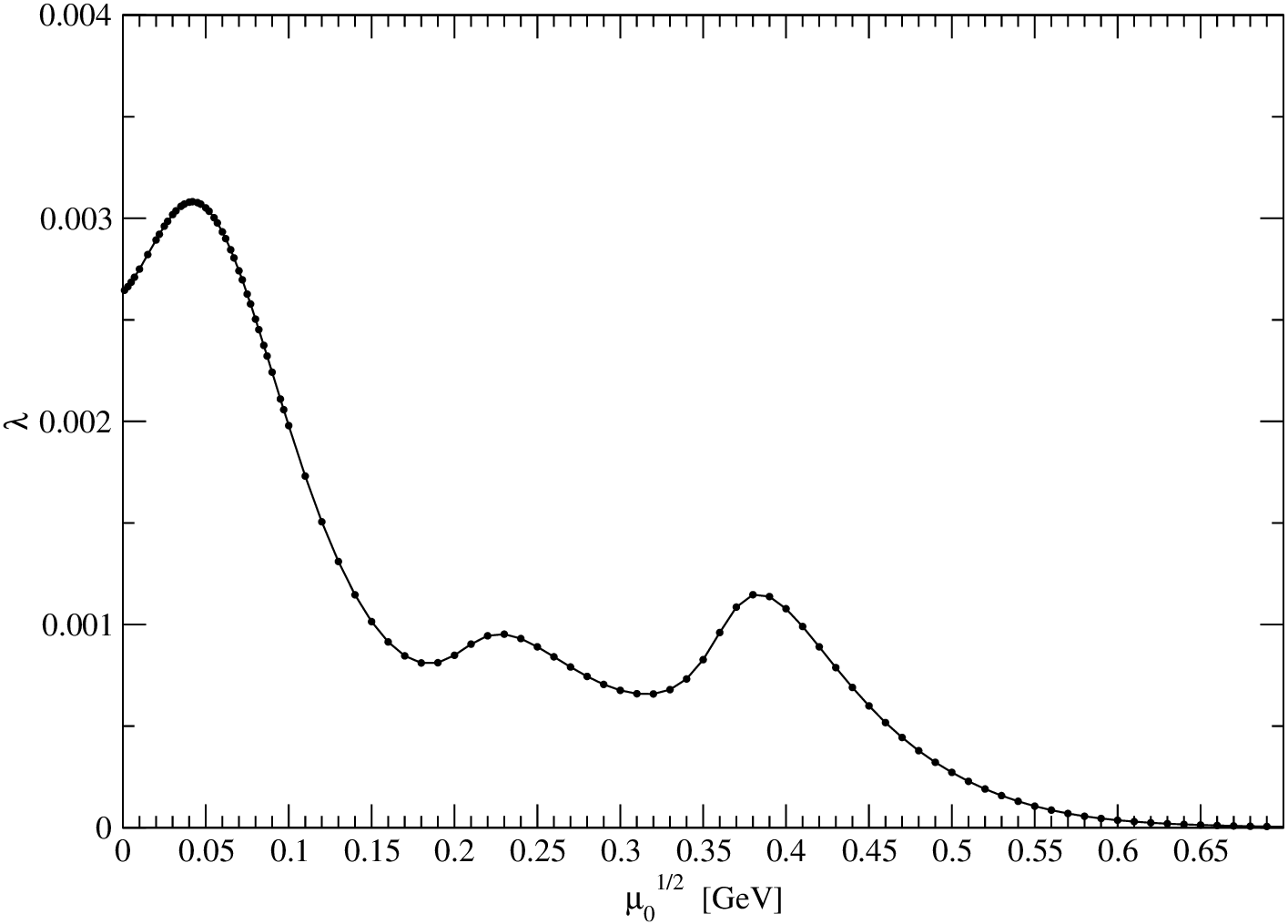} \quad
  \includegraphics[width=0.45\textwidth]{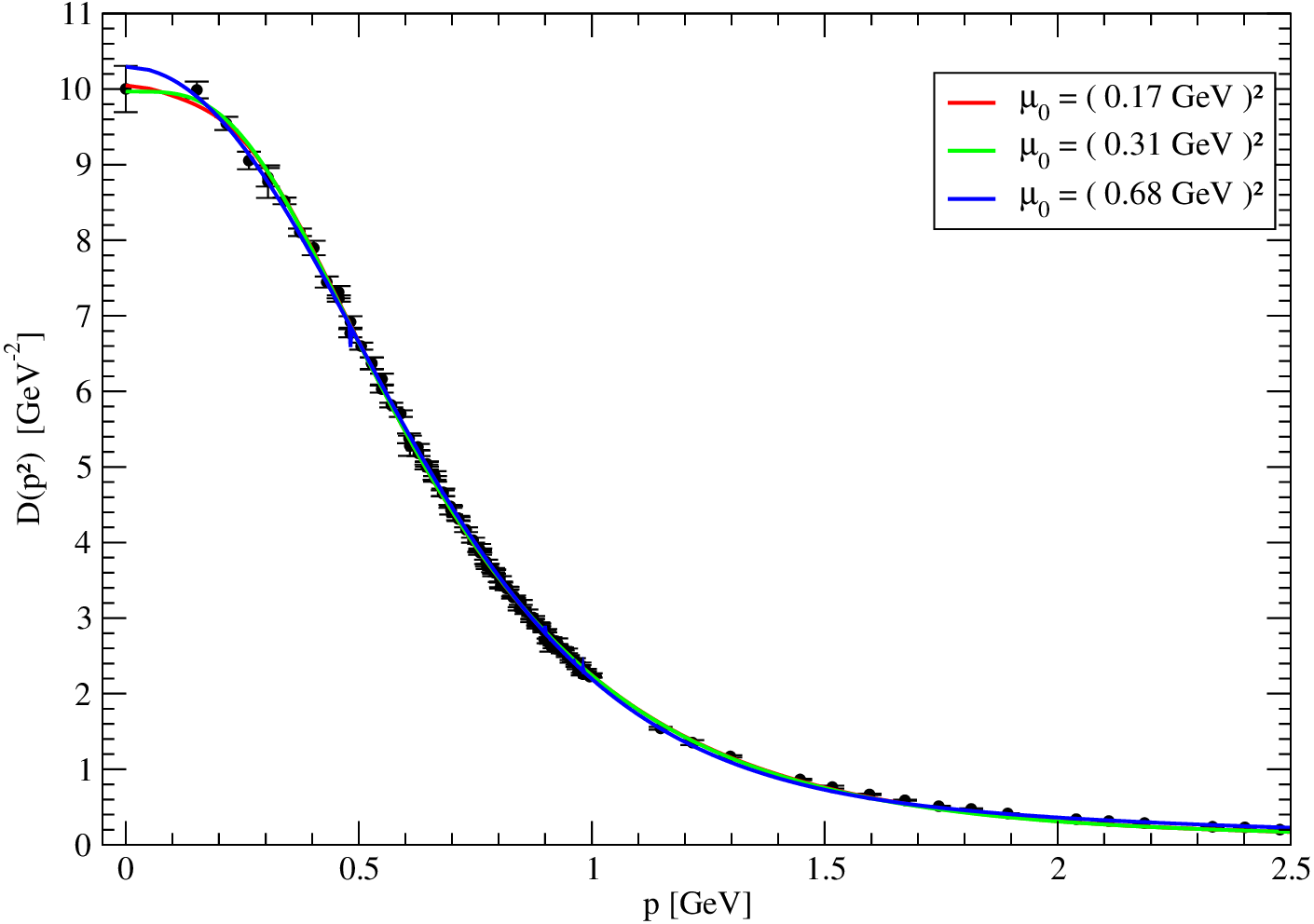}
\caption{The curve $\lambda ( \mu_0 )$ (left) together with the lattice gluon data ($80^4$ and $\beta = 6.0$ simulation)
and the rebuilt propagator for several optimal $\lambda$ parameters.}
\label{fig:4}       
\end{figure}

In order to invert the integral equation (\ref{Eq:spectral}) one needs to regularise it. Our calculation of $\rho ( \mu )$ uses
the Tikhonov regularisation, it follows~\cite{Dudal14} and it corresponds to minimising the following functional
\begin{equation}
 \mathcal{J}_\lambda =  \sum^{N}_{i = 1} \left[  \int^{+ \infty}_{\mu_0} d \mu ~ \frac{\rho ( \mu )}{p^2_i + \mu}  - D(p^2_i) \right]^2
 + \lambda ( \mu_0 )  \,  \int^{+ \infty}_{\mu_0} d \mu ~ \rho^2 ( \mu )
\end{equation}
where $N$ is the number of lattice momenta $p^2_i$ where the lattice propagator is defined, $\mu_0$ an infrared cutoff and $\lambda$
the regularisation parameter to be determined by the Morozov discrepancy principle,  which is a way to select the optimal $\lambda$ while keeping into account the quality of the data ($\sim$ total error margin). We refer to the literature for additional details. Afterwards, we choose the optimal $\lambda$'s at the minimum of the
curve $\lambda ( \mu_0 )$. Note that the minimisation procedure favours spectral functions with small L$_2 ( \mu_0 , + \infty )$ norms.
 This is open to improvement for which we refer to future work. The minimisation procedure reduces the problem to the solution of a linear system of equations. From its solution one can compute both the
spectral function and rebuild the propagator using (\ref{Eq:spectral}).

\begin{figure}
\centering
  \includegraphics[width=0.45\textwidth]{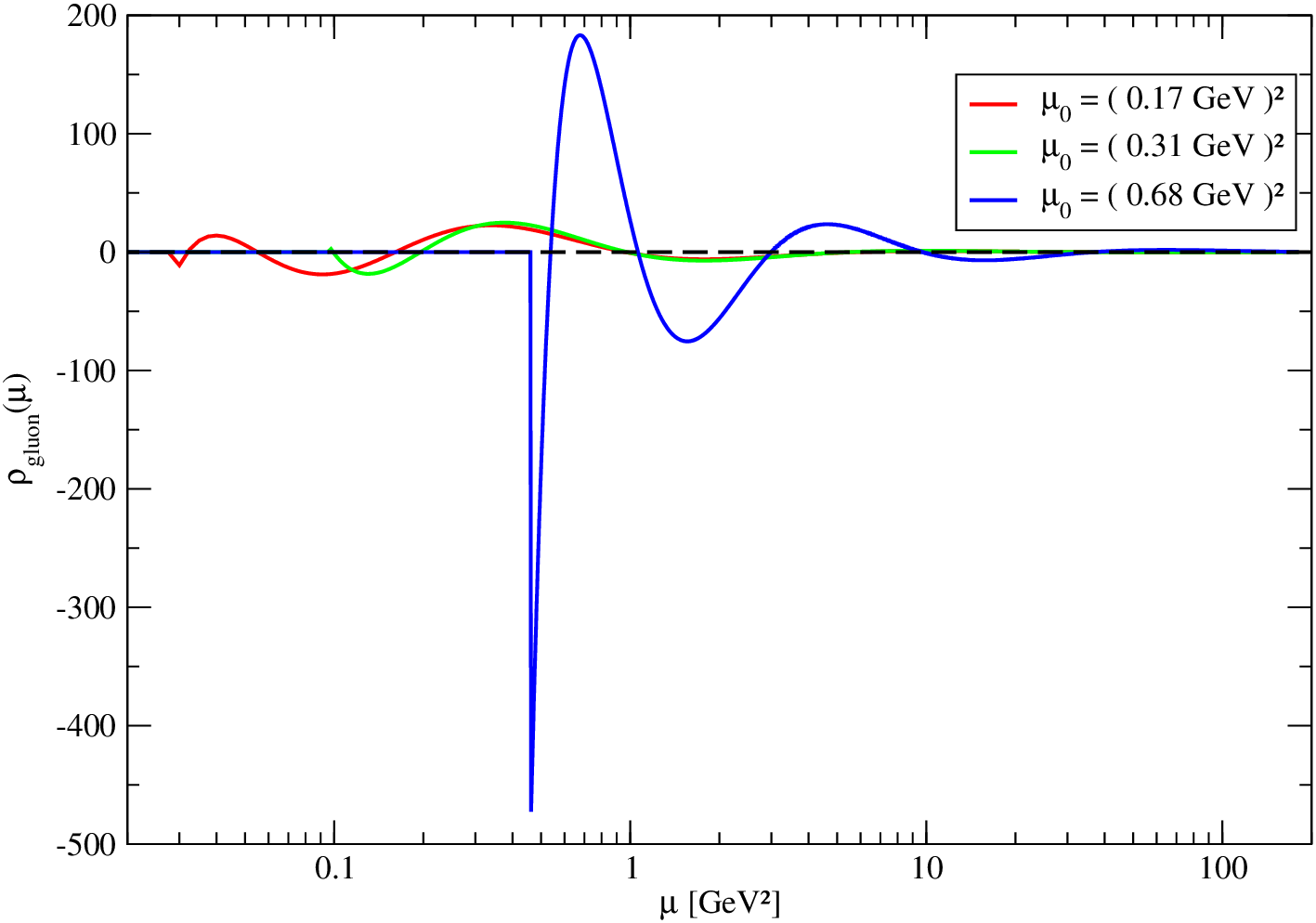} \quad
  \includegraphics[width=0.45\textwidth]{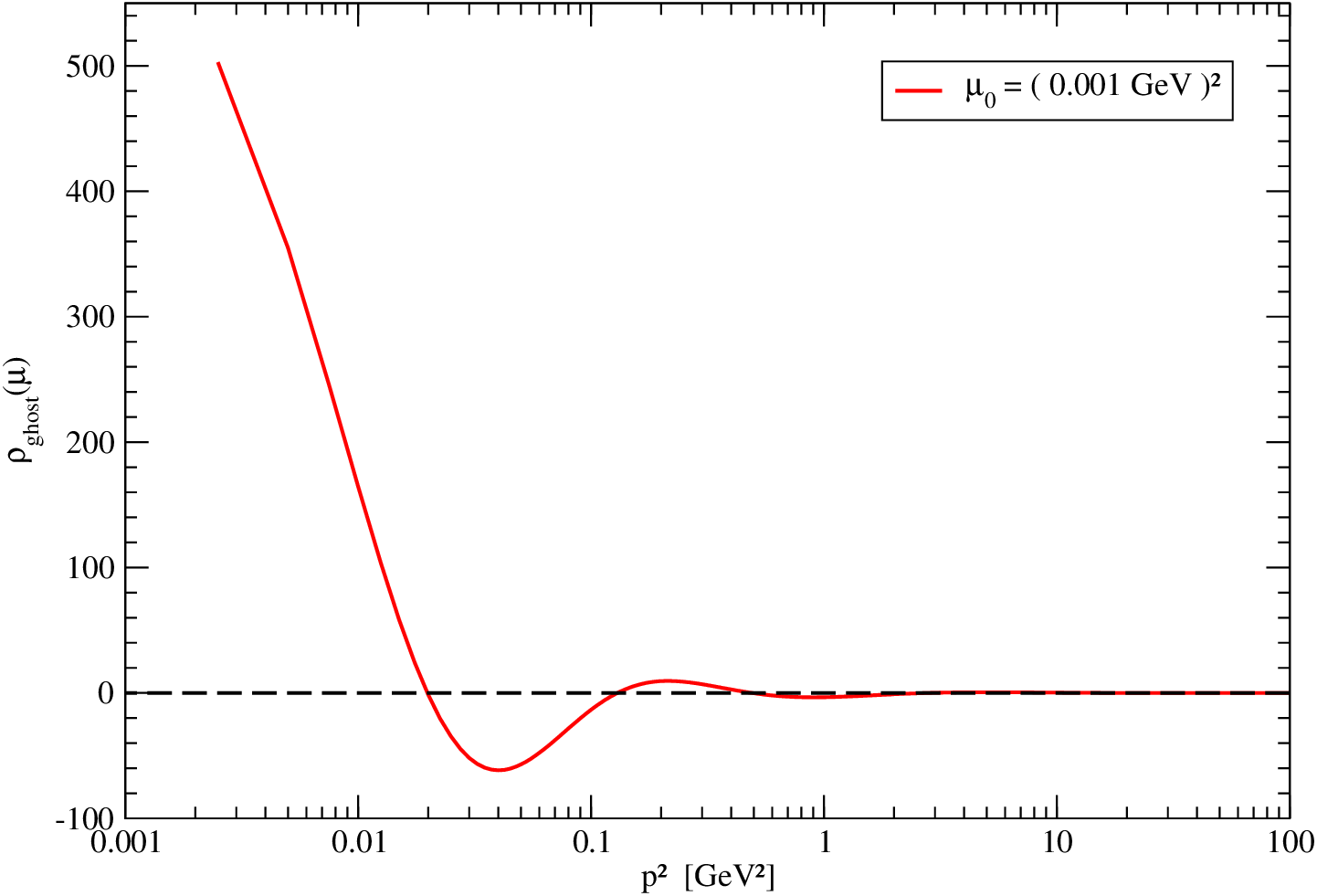}
\caption{The gluon (left) and ghost (right) spectral functions as foreseen by the Tikhonov regularisation and using the Morozov criterium.}
\label{fig:5}       
\end{figure}

From the simulation using the $80^4$ lattice and $\beta = 6.0$, the function $\lambda ( \mu_0 )$ together with the propagator data and the
rebuilt propagator for the various minima of the curve $\lambda ( \mu_0 )$ for the gluon propagator can be seen in Fig.~\ref{fig:4}.
The corresponding spectral function can be seen in Fig.~\ref{fig:5} and it clearly show regions of momenta where the sign of
$\rho$ is either positive or negative. According to our reasoning, this means that the Hilbert space of QCD does  not contain single
gluon states and, therefore, gluons have to be confined in composite states belonging to $\mathcal{H}$.

In Fig.~\ref{fig:4} we also report, for the first time, the preliminary result for the spectral function associated to the ghost propagator.
As can be seen, the lattice data
favours a small or vanishing infrared cutoff and, furthermore, $\rho ( \mu )$ shows a steepest increase as one approaches the zero momentum
suggesting that indeed $\rho ( \mu ) = \delta ( \mu ) + \cdots$, as one would expect from an essentially massless propagator. In this sense,
it seems that our inversion procedure is able to capture the main features of the spectral function.
 We will report on this in a forthcoming larger paper.

In what concerns the sum rule (\ref{Eq:sumrule}), our preliminary analysis suggests that indeed the spectral function $\rho ( \mu )$ computed using the
Tikhonov regularisation  is satisfied within the statistical precision of the simulation.
A more detailed study, including error analysis and continuum extrapolation is underway.
%

As a final comment we would like to point out that our estimate for $\rho  ( \mu )$ does not agree very well with that predicted by solving the Dyson-Schwinger
equations in the complex $p^2$ plane for the gluon and ghost propagators~\cite{Strauss12}.
Furthermore, the gluon propagator data is compatible with the tree level expression as predicted by the refined Gribov-Zwanziger expression \cite{Dudal:2008sp,Dudal:2010tf,Cucchieri:2011ig},
\begin{equation}
   D(p^2) = z \frac{ p^2 + m^2_1}{p^4 + m^2_2 \, p^2 + m^4_3} \ ,
\end{equation}
where $m_i$ are parameters with dimensions of mass, and, according to this expression, the gluon propagator has a pair of complex
conjugate poles at $p^2 \sim -0.3 \pm i \, 0.5$ GeV$^2$ for the simulations reported here. This result implies that the spectral
decomposition (\ref{Eq:spectral}) with a real $\rho ( \mu )$ does not hold for the gluon.  A more elaborate discussion of these issues will be discussed elsewhere, including possible generalizations of the K\"all\'{e}n-Lehmaan spectral integral.

For other recent works related to gluon, ghost and quark spectral functions at zero temperature, let us refer to \cite{Siringo:2016jrc,Rothkopf:2016luz}.

\section{The Three Gluon Vertex}

The three gluon one particle irreducible (1PI) Green function $\Gamma^{a_1 a_2 a_3}_{\mu_1 \mu_2 \mu_3} (p_1, p_2, p_3)$, our convention
being all  momenta are incoming which means that $p_1 + p_2 + p_3 = 0$, can be decomposed in terms of six Lorentz scalar form factors~\cite{Ball80}.
On the lattice one can not measure directly 1PI functions but, instead, one can access the complete Green function given by
\begin{equation}
   G^{(n)} \, ^{a_1 a_2 a_3}_{\mu_1 \mu_2 \mu_3} (p_1, p_2, p_3) = \langle A^{a_1}_{\mu_1} (p_1) A^{a_2}_{\mu_2} (p_2) A^{a_3}_{\mu_3} (p_3)
   \rangle \ ,
\end{equation}
which can be written in terms of propagators and the three gluon irreducible Green function
and, therefore, one can only measure combinations of the form factors appearing in $\Gamma$. Of the possible contractions and momenta
configurations, we will report results for
\begin{equation}
 G^{(n)} \, _{\alpha \mu \alpha} (p, 0,  -p) \, p_\mu = \mbox{Tr} \langle A_\alpha (p) \, A_{\mu} (0) A_{\alpha} (-p) \rangle \, p_\mu =
 \, \frac{N_c ( N^2_c - 1)}{4} \,  \left[ D(p^2) \right]^2 \, D(0) \, \Gamma (p^2) \, p^2
 \label{Eq:3pontos}
\end{equation}
where Tr stands for trace over color indices. This function was also investigated in the first lattice calculation of the three gluon vertex~\cite{Alles97}.
According to the analysis of the Dyson-Schwinger equations, one expects the form factor $\Gamma (p^2)$ to change sign for momenta
$\sim 200$ MeV. We recall that such a change of sign was already observed in the simulations of 3D SU(2) pure gauge theories~\cite{Cucchieri2008}
and recently also for 4D SU(3) pure Yang-Mills~\cite{Athenodorou2016,Duarte16b}.

\begin{figure}
\centering
  \includegraphics[width=0.45\textwidth]{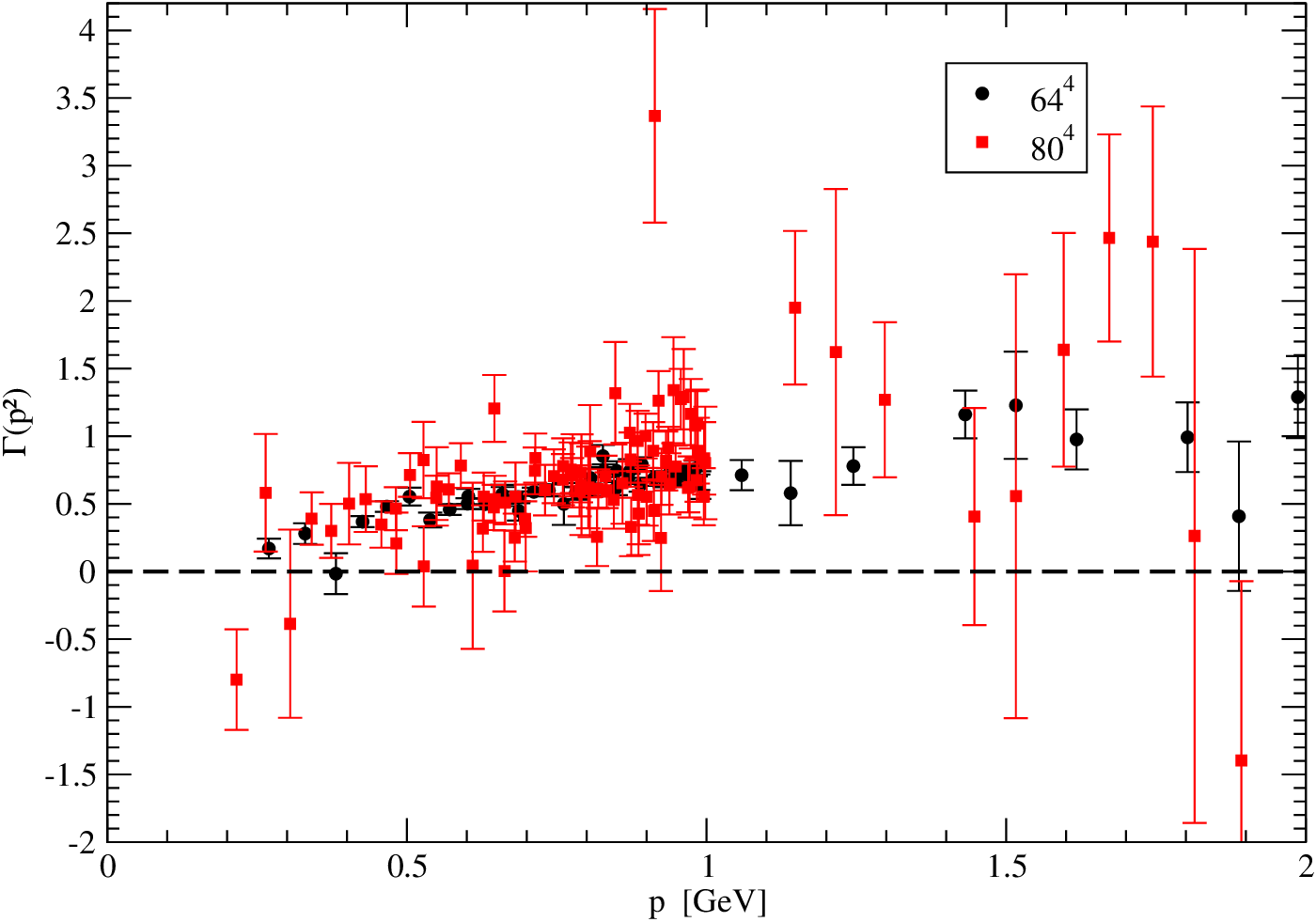} \quad
  \includegraphics[width=0.45\textwidth]{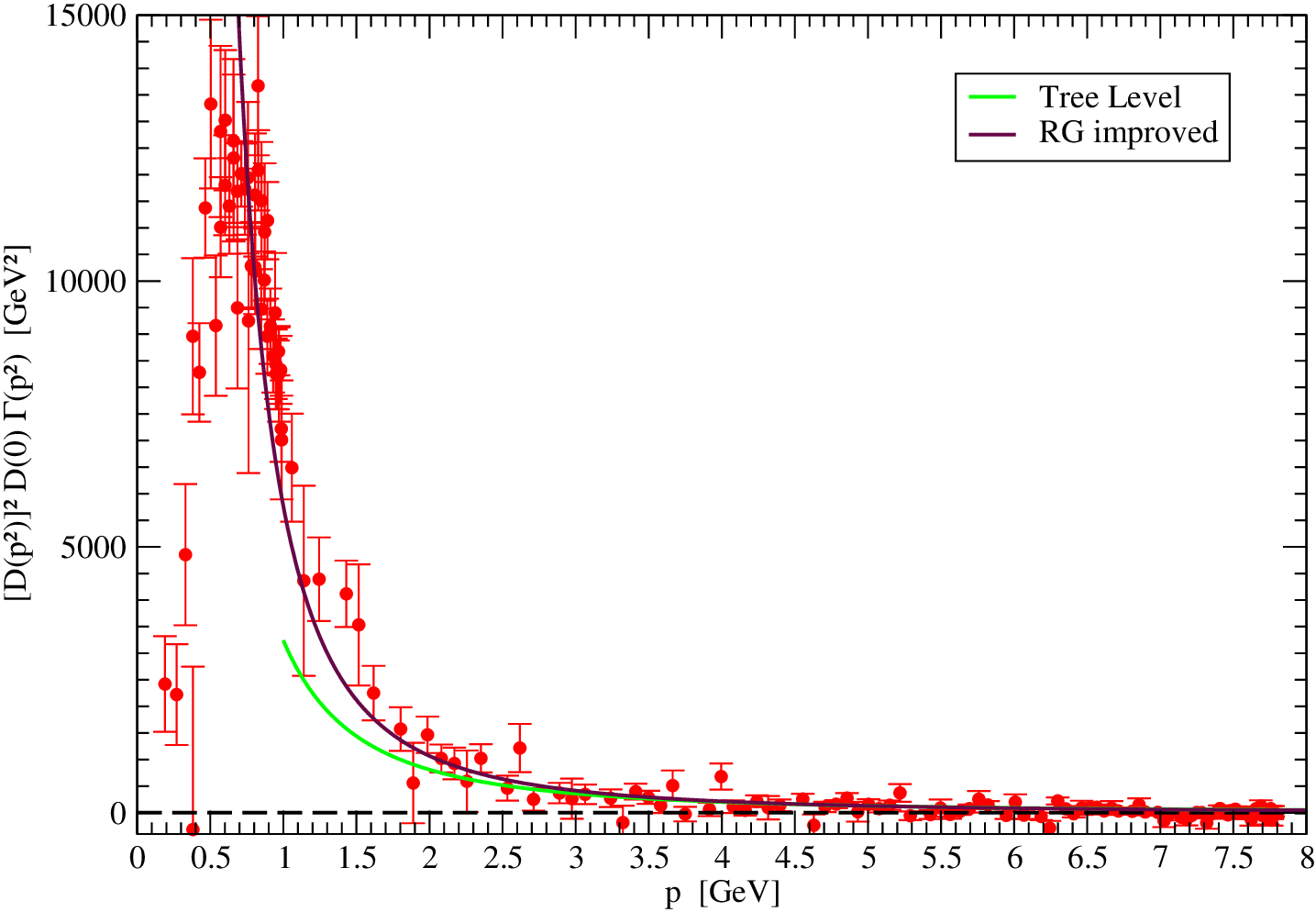}
\caption{The bare form factor associated to the complete Green function (\ref{Eq:3pontos}). The left plot refers to $\Gamma (p^2)$ up to
momenta 2 GeV. The right plot includes the propagator form factors together with the predictions of perturbation theory using $\Lambda_{QCD} = 200$ MeV.}
\label{fig:6}       
\end{figure}

For the computation of $\Gamma (p^2)$ we considered two sets of pure gauge ensembles both generated with the Wilson action and at
$\beta = 6.0$: (i) a smaller lattice $64^4$, which corresponds to a physical volume of (6.5 fm)$^4$, but with better statistics (2000 configurations);
(ii) a larger lattice of $80^4$, which has a physical volume of (8.1 fm)$^4$, but using less (279) configurations. As discussed in~\cite{Duarte16b}, if
at the level of the gluon propagator we see no difference for different types of momenta, for the three point function, due to the breaking of
rotational symmetry, the results depend on the type of momenta considered. In order to reduce possible effects associated to the breaking of
the group $O(4)$, we report results only for momenta of types $(n_x, n_y, 0, 0)$, $(n_x, n_y, n_z, 0)$ and $(n_x, n_y, n_z, n_t)$.

The form factors associated to $G^{(n)}$ can be seen in Fig.~\ref{fig:6}. On the left hand side, $\Gamma (p^2)$ is plotted for momenta below 2 GeV and
the data suggest that this form factors changes sign for $p \sim 200$ MeV as expected from the analysis made in \cite{Binosi13,eichmann14}.
A more careful analysis of the lattice data suggests that the zero crossing for 4D SU(3) Yang-Mills theory should take place in the
interval 216 - 264 MeV in good agreement also with the results for 3D SU(3) Yang-Mills simulations \cite{Cucchieri2008}.

On the right side plot of Fig.~\ref{fig:6}, we display the results of our simulation for the $80^4$ over the full range of momenta, together with
the predictions of perturbation theory. As the figure shows, at sufficiently high momenta the results of the lattice simulation are in good agreement
with renormalisation group improved perturbation theory.

\section{Summary and Conclusions}
In this work we report the results of simulating pure Yang-Mills theory on a finite lattice. Our simulation confirms that the gluon propagator
is finite and non-vanishing at all momenta and that it is the propagator of a massive-like  boson. On the other hand, the ghost reproduces the
propagator of an essentially massless degree of freedom. Further, the simulations show that for the propagators the
finite volume effects are under control for lattice volumes of (6.5 fm)$^4$ and above and that the most important
lattice artifacts are related to the value of the lattice spacing. While the gluon propagator seems to approach its continuum value from below, the ghost propagator seems to approach its continuum limit from above.

We have also investigated the gluon and ghost spectral representations. The two spectral functions have positive and negative values for  different
ranges of momenta, meaning that the spectrum of QCD does not account for single gluon (and evidently) ghost states.
Gluon and ghost only show up in the eigenstates of $\mathcal{H}$ as constituents of other particles and, in this sense, they are confined particles.
For what concerns the spectral functions, there are still a number of problems that need to be solved, in particular in relation to the ultraviolet tails
which are no well reproduced by the method used in the current work. This can be understood, at least partially, due to the large number of data
in the low momenta region and relatively scarce number of data points for momenta above two or three GeV. Ways to overcome these problems are
currently under investigation.

The good agreement between lattice simulations and theoretical analyses (refined Gribov-Zwanziger, Curci-Ferrari models, Dyson-Schwinger equations,
functional renormalization group equations, etc.)
for the propagators allows us to conclude that we have now a good theoretical picture for the glue and ghost dynamics.

For what concerns the investigation of the three gluon vertex, our simulations support that a change of sign for the form factor $\Gamma$ takes
place at momenta $\sim 200$ MeV. The analysis of the simulation shows that the artifacts due to the breaking of rotational symmetry are larger for
the three point function than for the two point functions. The simulations also show that in the high momenta regime they agree well with the results
from perturbation theory. We aim to proceed our investigation of the three point function in order to describe the form factors that are associated
to the three gluon vertex and we hope that our results will also open the road to study gluon correlation functions with a
 higher number of external legs.

%
%
%

\begin{acknowledgements}
O.~Oliveira, A.~G.~Duarte and P.~J.~Silva acknowledge financial support from FCT Portugal under contract with reference UID/FIS/04564/2016.
P.~J.~Silva acknowledge supported by FCT under Contracts No. SFRH/BPD/40998/2007 and SFRH/BPD/109971/2015.
The computing time was provided by the Laboratory for Advanced Computing at the University
of Coimbra and by PRACE projects COIMBRALATT (DECI-9) and COIMBRALATT2
(DECI-12).
\end{acknowledgements}



\end{document}